\documentclass[twocolumn,showpacs,aps,prl,letterpaper]{revtex4}
\usepackage{graphicx}
\usepackage{dcolumn}
\usepackage{amsmath}
\usepackage{epsfig}
\usepackage{hyperref}

\long\def\inst#1{\par\nobreak\kern 4pt\nobreak
    {\itshape #1}\par\vskip 10pt plus 3pt minus 3pt}
\RequirePackage{xspace}
\usepackage{relsize}

\newcommand{\PreserveBackslash}[1]{\let\temp=\\#1\let\\=\temp}
\newcolumntype{C}[1]{>{\PreserveBackslash\centering}p{#1}}
\newcolumntype{R}[1]{>{\PreserveBackslash\raggedleft}p{#1}}
\newcolumntype{L}[1]{>{\PreserveBackslash\raggedright}p{#1}}

\begin{document}

\title{\large \bfseries \boldmath Spectra of heavy-light mesons in a relativistic  model}
\author{Jing-Bin Liu} \email{liujb@ihep.ac.cn}
\author{Cai-Dian L$\ddot{\rm u}$} \email{lucd@ihep.ac.cn}
\affiliation{Institute of High Energy Physics, Beijing 100049, People's Republic of China}

\date{October 18, 2016}

\begin{abstract}
The spectra and wave functions of heavy-light mesons are calculated within a relativistic quark model, which is based on a heavy-quark expansion of the instantaneous Bethe-Salpeter equation by applying the Foldy-Wouthuysen transformation. The kernel we choose is the standard combination of linear scalar and Coulombic vector. The effective Hamiltonian for heavy-light quark-antiquark system is calculated up to order $1/m_Q^2$. Our results are in good agreement with available experimental data except for the anomalous $D_{s0}^*(2317)$ and $D_{s1}(2460)$ states. The newly observed heavy-light meson states can be accommodated successfully in the relativistic quark model with their assignments presented. The $D_{sJ}^*(2860)$ can be interpreted as the $|1^{3/2}D_1\rangle$ and $|1^{5/2}D_3\rangle$ states being members of the 1D family with $J^P=1^-$ and $3^-$.
\end{abstract}


\maketitle

\section*{I Introduction}

Great experimental  progress has been achieved in studying the spectroscopy of heavy-light mesons in the last decades \cite{D2010,LHCb,Ds2632,Ds2860,Ds2700,Ds2009,LHCbspin13b,LHCbspin13a}. In the charm sector, several new excited charmed meson states were discovered in addition to the low-lying states. For $D_J$ mesons, the excited resonances $D(2740)^0$, $D^*(2760)$ \cite{D2010}, $D_J(2580)^0$, $D^*(2650)$ and $D^*(3000)$ \cite{LHCb} were found in the $D^{(*)}\pi$ invariant mass spectrum by the BaBar and LHCb Collaborations. While for $D_{sJ}$ mesons, besides the well-established $1S$ and $1P$ charmed-strange states, the excited resonances $D_{sJ}(2632)$ \cite{Ds2632}, $D_{sJ}(2860)$ \cite{Ds2860}, $D_{sJ}(2700)$ \cite{Ds2700} and $D_{sJ}(3040)$ \cite{Ds2009} were observed in the $D^{(*)}K$ invariant mass distribution by the two collaborations. In the b-flavored meson sector, several excited states were studied in experiment as well as the ground  $B$ and $B_s$ meson states \cite{PDG}. The strangeless resonances $B_J(5840)^0$ and $B(5970)^0$ were found in the $B\pi$ invariant mass spectrum by the LHCb and CDF Collaborations, respectively \cite{LHCba,CDFa}. While the stranged $B^*_{sJ}(5850)$ were observed in the $B^{(*)}K$ invariant mass distribution by the OPAL Collaboration \cite{OPAL}.

The heavy-light meson spectroscopy plays an important role in understanding the strong interactions between quark and antiquark. In the meanwhile it provides a powerful test of the various phenomenological quark models inspired by QCD.
Heavy-light mesons have been investigated extensively in relativistic quark models \cite{GI,cnp,cccn,cnp1,cflns,ymz,GMS}, where many relativistic potential models are constructed by modifying or relativizing nonrelativistic quark potential models and additional phenomenological parameters are employed. For heavy-light system, one needs a model that can retain the relativistic effects of the light quark. In this work we resort to the originally relativistic Bethe-Salpeter equation \cite{BS}. The Bethe-Salpeter approach were widely used in studying mesons so as to embody the relativistic dynamics \cite{LS1,LS2,FK1,FK2,WWFJ,CCh}. While it is rather difficult to solve the Bethe-Salpeter equation for meson states, especially when considering states with large angular momentum quantum number. In order to study the spectra of heavy-light mesons systematically, we choose to reduce the Bethe-Salpeter equation in the first place.

In our previous work \cite{ly3}, we apply the instantaneous approximation and obtain an equation equivalent to the Bethe-Salpeter equation. The Hamiltonian for heavy-light quark-antiquark system is expanded to order $1/m_Q$ by applying Foldy-Wouthuysen transformation to the equivalent equation. We find that the leading Hamiltonian is actually not Dirac-like. The interaction we derive is essentially different from the Breit interaction \cite{MKM,TM,MM}. In this paper we extend and improve our study on the spectra of the heavy-light mesons $D$, $D_s$, $B$ and $B_s$. The running of the coupling constant is considered. Moreover, the $1/m_Q^2$ correction is calculated. Many papers only considered the leading $1/m_Q$ term in the heavy-quark expansion \cite{ly3,PE,JZ,DJ,DHJ}. Our calculation shows that the $1/m_Q^2$ corrections to the masses of the mesons are around 50 MeV, which is too large to be neglected.
The parameters in the equations are determined by fitting the masses of the $1S$ and $1P$ meson states presented by Particle Data Group (PDG) \cite{PDG}, while the states beyond $1P$ are calculated as a prediction. We find that in the Bethe-Salpeter formalism the linear confining parameter, i.e. the string tension, actually depend on the masses of the constituent quark and antiquark in mesons. The large discrepancy between experimental data and our previous work is decreased in this work. The newly  observed heavy-light meson states can be accommodated successfully in our predicted spectra.

This paper is organized as follows. In the next section, we have a brief review of the relativistic quark model. Section III is for the solution of the wave equation and the peturbative corrections. In Section IV, we have numerical results and discussions. The last section is for a brief summary.

\section*{II The model}

According to the conventional constituent-quark model, the mesons can be seen as a composition of a quark and an antiquark. In the Bethe-Salpeter formalism, the eigenequation for quark-antiquark systems has the general form \cite{BS}:
\begin{eqnarray}
&&(p\!\!\!/_1-m_1)\chi(p_1, p_2)(p\!\!\!/_2+m_2)\nonumber\\
&&=\frac{1}{(2\pi)^4} \int\!d^4p'_1 d^4p'_2 \overline K(p_1, p_2; p'_1, p'_2)\chi(p'_1, p'_2),   \label{e1}
\end{eqnarray}
where $p_1$ and $p_2$ relate to the total momentum $P$ and the relative momentum $p$ as follows:
\begin{eqnarray}
p_1&=&\alpha_1 P-p, \;\; \alpha_1=\frac{m_1}{m_1+m_2},\\
p_2&=&\alpha_2 P+p, \;\; \alpha_2=\frac{m_2}{m_1+m_2}.
\end{eqnarray}
Using the energy-momentum conservation, i.e. $p'_1+p'_2=p_1+p_2$, the Eq.(\ref{e1}) can be simplified as:
\begin{equation}
(p\!\!\!/_1-m_1)\chi(p,P)(p\!\!\!/_2+m_2)=\int\!\frac{d^4p^\prime}{(2\pi)^4}\overline K(p,p^\prime,P)\chi(p^\prime,P).   \label{e2}
\end{equation}

Here we choose the interaction kernel as the standard Coulomb-plus-linear form, which is one-gluon-exchange (OGE) dominant at short distances with linear confinement at long distances. If one applies the instantaneous approximation, i.e. neglecting the frequency dependence, the kernel can be written as:
\begin{equation}
\overline K(p,p^\prime,P)=\gamma^{(1)}\cdot\gamma^{(2)}V_v(-\boldsymbol{k}^2)+V_s(-\boldsymbol{k}^2),  \label{e5}
\end{equation}
where the transferred momentum $\boldsymbol{k}$ is defined as:
\begin{equation}
\boldsymbol{k}=\boldsymbol{p}-\boldsymbol{p}^\prime.
\end{equation}

Since the interaction kernel $\overline K(\boldsymbol{p},\boldsymbol{p}^\prime,P)$ is no longer dependent on $p^{\prime0}$, we can perform the integration over  $p^{\prime0}$ in Eq. (\ref{e2}). After transforming the instantaneous Bethe-Salpeter equation into coordinate space, the wave function of the eigenequation decouples from time coordinate \cite{GR,CCW,CC}.

In our previous work \cite{ly3}, with the help of projection operators for the wave function we find that the instantaneous Bethe-Salpeter equation is equivalent to the following equation:
\begin{equation}
\left(\omega_1+\omega_2+\frac{1}{2}(h_1+h_2)U(\boldsymbol{r})\frac{1}{2}(h_1+h_2)-h_1E\right)\phi(\boldsymbol{r})=0,\label{e7}
\end{equation}
where the superscript ``1" and ``2" stand for the heavy quark $Q$ and the light antiquark $\bar{q}$ in the $Q\bar{q}$ meson, respectively. The operators in the above equation are defined as:
\begin{eqnarray}
\omega_i(\boldsymbol{p})&=&\sqrt{\boldsymbol{p}^2+m_i^2},\;\; i=1,2 \label{e8} \\
h_i(\boldsymbol{p})&=&\frac{H_i(\boldsymbol{p})}{\omega_i(\boldsymbol{p})},\;\; i=1,2\label{e9}
\end{eqnarray}
with the free Dirac Hamiltonians
\begin{eqnarray}
H_1(\boldsymbol{p})&=&{\beta}^{(1)}m_1-{\boldsymbol{\alpha}}^{(1)}\cdot\boldsymbol{p},\label{e10}\\
H_2(\boldsymbol{p})&=&{\beta}^{(2)}m_2+{\boldsymbol{\alpha}}^{(2)}\cdot\boldsymbol{p}.\label{e11}
\end{eqnarray}
Inserting Eqs. (\ref{e10}) and (\ref{e11}) into Eq.(\ref{e9}), we can verify the relation:
\begin{equation}
h_i^2(\boldsymbol{p})=1,\;\; i=1,2.\\\label{z12}
\end{equation}

The interaction potential $U(\boldsymbol{r})$ in Eq. (\ref{e7}) is directly derived from the instantaneous Bethe-Salpeter equation and closely related to the interaction form we assumed in the kernel. It can be written as:
\begin{equation}
U(\boldsymbol{r})=U_1(\boldsymbol{r})+U_2(\boldsymbol{r})
\end{equation}
with
\begin{eqnarray}
U_1(\boldsymbol{r})&=&\beta^{(1)}\beta^{(2)}V_s(r)+V_v(r),\\
U_2(\boldsymbol{r})&=&-\frac{1}{2}[\,{\boldsymbol{\alpha}}^{(1)}\cdot{\boldsymbol{\alpha}}^
{(2)}+({\boldsymbol{\alpha}}^{(1)}\cdot\hat{\boldsymbol{r}})
({\boldsymbol{\alpha}}^{(2)}\cdot\hat{\boldsymbol{r}})\,]V_v(r),\nonumber\\
\end{eqnarray}
where $V(r)$ and $V(-\boldsymbol{k}^2)$ are related to each other according to Fourier transformation.

The instantaneous Bethe-Salpeter equation as an integral equation now is equivalent to a less complicated differential equation shown in Eq. (\ref{e7}), but still it is difficult to solve.
For heavy-light systems, the heavy quark effective theory is applied. It is reasonable to consider the heavy-quark expansion, i.e. the $1/m_Q$ expansion. One can reduce the equivalent eigenequation by calculating the interactions of the heavy-light quark-antiquark meson order by order.

Our goal can be achieved by employing the Foldy-Wouthuysen transformation \cite{bGR}. The operators involved in Eq. (\ref{e7}) can be divided into two sets: the ``odd" $\mathcal{O}$ and the ``even" $\mathcal{E}$. The name ``odd" denotes that the operators couple the large and small components of the Dirac spinor, while the ``even" operators are diagonal with respect to the large and small components. The main idea of the Foldy-Wouthuysen transformation is to apply a unitary transformation $U$ which retains the ``even" operators and eliminates the ``odd" operators.
If one writes the original Hamiltonian as:
\begin{equation}
H=\beta \, m+\mathcal{E}+\mathcal{O}, \label{e16}
\end{equation}
according to Foldy and Wouthuysen, one obtains the transformed Hamiltonian:
\begin{eqnarray}
\tilde{H}&=&U^{-1}H\, U\nonumber\\
&=&\beta m+\mathcal{E}+\frac{\beta}{2m}\mathcal{O}^2+\frac{1}{8m^2}[\, [ \mathcal{O},\,\mathcal{E} ],\,
\mathcal{O}\, ]+\cdots  \label{e17}
\end{eqnarray}

The reduction by performing the Foldy-Wouthuysen transformation on Eq. (\ref{e7}) is detailed in our previous work \cite{ly3}. Instead of $\beta m$ being the main term in the common Dirac Hamiltonian shown in Eq. (\ref{e16}), the dominant term is $\beta E$ in our case:
\begin{eqnarray}
-h_1 E&=&\frac{{\boldsymbol{\alpha}}^{(1)}\cdot\boldsymbol{p}-{\beta}^{(1)}m_1}{\omega_1}E \nonumber \\
&=&-\beta^{(1)} E+\frac{\boldsymbol{\alpha}^{(1)}\cdot\boldsymbol{p} }{\omega_1}E -\beta^{(1)}\left(\frac{m_1}{\omega_1}-1\right)E. \nonumber \\ \label{e18}
\end{eqnarray}

The reduction result is calculated to order $1/m_Q$ in our previous work \cite{ly3}. With the similar procedure, here we extend the result  to order $1/m_Q^2$. By inserting the ``odd" and ``even" operators of Eq. (\ref{e7}) into Eq. (\ref{e17}), we obtain the Hamiltonian expansion. After the Foldy-Wouthuysen transformation, we have:
\begin{eqnarray}
\tilde{H}&=&\tilde{H}_0+\tilde{H}^{\prime}
\label{e215}
\end{eqnarray}
with
\begin{eqnarray}
\tilde{H}_0&=&\omega_1+\omega_2
+\frac{1}{2}\left(1+h_2\right)U_1\frac{1}{2}\left(1+h_2\right),\label{e212}
\end{eqnarray}
the perturbative term $\tilde{H}^{\prime}$ consists of various terms of order $1/m_Q$ and $1/m^2_Q$, we divide it into three parts:
\begin{eqnarray}
\tilde{H}^{\prime}=\tilde{H}_1^{\prime}+\tilde{H}_a^{\prime}+\tilde{H}_b^{\prime},
\end{eqnarray}
where
\begin{eqnarray}
\tilde{H}_1^{\prime}&=&\frac{1}{2}\left(1+h_2\right)\{-\frac{{\boldsymbol{\alpha}}^{(1)}\cdot\boldsymbol{p}}{2 m_1},\;U_2\}\frac{1}{2}\left(1+h_2\right), \label{e212a} \\
\tilde{H}_a^{\prime}&=&\frac{1}{2}\left(1+h_2\right)\frac{{\boldsymbol{\alpha}}^{(1)}\cdot\boldsymbol{p}}{2m_1}U_1
\frac{{\boldsymbol{\alpha}}^{(1)}\cdot\boldsymbol{p}}{2m_1}\frac{1}{2}\left(1+h_2\right)\nonumber\\
&-& \frac{1}{2}\left(-3+h_2\right)\frac{{\boldsymbol{p}}^2}{8m_1^2} U_1 \frac{1}{2}\left(1+h_2\right)
+h.c.    , \label{e212b} \\
\tilde{H}_b^{\prime}&=&\frac{1}{2}\left(1+h_2\right)U_2\frac{1}{2}(1+\beta^{(1)}h_2)U_1\frac{{\boldsymbol{\alpha}}^{(1)}\cdot\boldsymbol{p}}{4Em_1}+h.c. \nonumber\\
&+&\frac{1}{2}\left(1+h_2\right)U_2\frac{1}{2}(\beta^{(1)}+h_2)\frac{{\boldsymbol{\alpha}}^{(1)}\cdot\boldsymbol{p}}{4Em_1}U_1\frac{1}{2}\left(1+h_2\right)
\nonumber\\ &+& h.c. \label{e212}
\end{eqnarray}

We can simplify the above equations by inserting an identity matrix $(\gamma^{(1)}_5)^2=1$ between two odd operators of the heavy quark, with the help of the relations $\{\gamma_5,\;\beta\}=0$, $[\gamma_5,\;\boldsymbol{\alpha}]=0$, $\gamma_5\boldsymbol{\alpha}=\boldsymbol{\Sigma}$ and the substitutions $\beta^{(1)}\rightarrow1$, $\boldsymbol{\Sigma}^{(1)}\rightarrow\boldsymbol{\sigma}^{(1)}$. Moreover, we can take the substitution $h_2\rightarrow1$ if $h_2$ appears at the ends of the expression of $\tilde{H}^{\prime}$ as in Eqs. (\ref{e212a}$\sim$\ref{e212}), since the corrections of $\tilde{H}^{\prime}$ are calculated as a perturbation to $\tilde{H}_0^{\prime}$.

With the considerations above, we obtain our final Hamiltonian
\begin{eqnarray}
H&=&H_0+H^{\prime},
\end{eqnarray}
where the leading order Hamiltonian $H_0$ has the form
\begin{eqnarray}
H_0&=&\omega_1+\omega_2
+\frac{1}{2}\left(1+h_2\right)\overline{U}_1\frac{1}{2}\left(1+h_2\right), \label{z14}
\end{eqnarray}
and the subleading Hamiltonian $H^{\prime}$ to order $1/m^2_Q$ can be written as:
\begin{eqnarray}
H^{\prime}=H_1^{\prime}+H_a^{\prime}+H_b^{\prime}, \label{e27}
\end{eqnarray}
with
\begin{eqnarray}
H_1^{\prime}&=&-\frac{1}{2}\{\frac{{\boldsymbol{\sigma}}^{(1)}\cdot\boldsymbol{p}}{m_1},\;\widetilde{U}_2\},\label{z9}\\
H_a^{\prime}&=&\frac{1}{4}\frac{{\boldsymbol{\sigma}}^{(1)}\cdot\boldsymbol{p}}{m_1}\widetilde{U}_1\frac{{\boldsymbol{\sigma}}^{(1)}\cdot\boldsymbol{p}}{m_1}
+\frac{1}{8}\{\frac{{\boldsymbol{p}}^2}{m_1},\;\overline{U}_1\},\label{z10}\\
H_b^{\prime}&=&\frac{1}{4E}\left(\widetilde{U}_2\frac{1}{2}(1-h_2)\widetilde{U}_1\frac{{\boldsymbol{\sigma}}^{(1)}\cdot\boldsymbol{p}}{m_1}+h.c.\right) \nonumber\\
&-&\frac{1}{4E}\left(\widetilde{U}_2\frac{1}{2}(1-h_2)\frac{{\boldsymbol{\sigma}}^{(1)}\cdot\boldsymbol{p}}{m_1}\overline{U}_1+h.c.\right),\label{z11}
\end{eqnarray}
the interaction potentials $\overline{U}_1(\boldsymbol{r})$, $\widetilde{U}_1(\boldsymbol{r})$ and $\widetilde{U}_2(\boldsymbol{r})$ in the above equations are defined as:
\begin{eqnarray}
\overline{U}_1(\boldsymbol{r})&=&V_v(r)+\beta^{(2)}V_s(r),\\
\widetilde{U}_1(\boldsymbol{r})&=&V_v(r)-\beta^{(2)}V_s(r),\\
\widetilde{U}_2(\boldsymbol{r})&=&-\frac{1}{2}[{\boldsymbol{\sigma}}^{(1)}\cdot{\boldsymbol{\alpha}}^
{(2)}+({\boldsymbol{\sigma}}^{(1)}\cdot\hat{\boldsymbol{r}})
({\boldsymbol{\alpha}}^{(2)}\cdot\hat{\boldsymbol{r}})]V_v(r).\nonumber\\
\end{eqnarray}

The leading order Hamiltonian $H_0$ we obtain for heavy-light quark-antiquark system in Eq. (\ref{z14}) is not Dirac-like as in Refs. \cite{JZ,EFG}. Its form is more like the form used in relativized quark models \cite{GI,LY1,LY2}. As for double-heavy system, we have $h_2\rightarrow1$ and $\beta^{(2)}\rightarrow1$, then Eq. (\ref{z14}) can be reduced to:
\begin{equation}
H_0^{\rm{Schr}}=\omega_1+\omega_2+V_v(r)+V_s(r), \label{b39}
\end{equation}
which is the Schr\"{o}dinger formalism extensively used in nonrelativistic or semirelativistic quark models.

\section*{III Solution of the Wave Equation}

In this section, we solve the eigenequation of the leading order Hamiltonian $H_0$ in Eq. (\ref{z14}). Before doing this, we would like to discuss the properties of the solution of the eigenequation associated with $H_0$.

The eigenequation of $H_0$ can be written as:
\begin{equation}
\left(\omega_1+\omega_2
+\frac{1}{2}\left(1+h_2\right)\overline{U}_1\frac{1}{2}\left(1+h_2\right)-E\right)\psi=0,\label{z1}
\end{equation}
the above equation is equivalent to:
\begin{equation}
h_2\left(\omega_1+\omega_2
+\frac{1}{2}\left(1+h_2\right)\overline{U}_1\frac{1}{2}\left(1+h_2\right)-E\right)\psi=0,\label{z2}
\end{equation}
which is equivalent to:
\begin{equation}
\left(\omega_1+\omega_2
+\frac{1}{2}\left(1+h_2\right)\overline{U}_1\frac{1}{2}\left(1+h_2\right)-E\right)h_2\psi=0.\label{z3}
\end{equation}
From Eqs. (\ref{z1}) and (\ref{z3}), we have:
\begin{equation}
h_2\psi=c\psi,
\end{equation}
and since $(h_2)^2=1$,
\begin{equation}
c=\pm1.
\end{equation}
When we take $c=-1$, Eq. (\ref{z1}) is transformed to:
\begin{equation}
(\omega_1+\omega_2-E)\psi=0,\label{z4}
\end{equation}
which is not the correct eigenequation for bound systems we are interested in. Thus we have only $c=+1$. This is the reason for the substitution  $h_2\rightarrow1$ we use in the last section.

While if all the eigenfunctions of $H_0$ for bound states satisfy the relation
\begin{equation}
h_2\psi=\psi, \label{z5}
\end{equation}
the eigenfunction set of $H_0$ is NOT complete.

An complete set is needed to construct the identity operator $1=\sum_{i}|\psi_i><\psi_i|$ in order to calculate the perturbative correction of $H_b^{\prime}$, thus we construct a new Hamiltonian. Inspired by the relation $h_2\psi=\psi$, we transform the potential term in Eq.(\ref{z1}) as:
\begin{eqnarray}
&&\frac{1}{2}\left(1+h_2\right)\overline{U}_1\frac{1}{2}\left(1+h_2\right)\nonumber\\
&&=\frac{1}{4}(\overline{U}_1+h_2\overline{U}_1+\overline{U}_1h_2+h_2\overline{U}_1h_2)\nonumber\\
&&\Rightarrow\frac{1}{4}(\overline{U}_1h_2+h_2\overline{U}_1+\overline{U}_1h_2+h_2\overline{U}_1)\nonumber\\
&&=\frac{1}{2}\{h_2,\;\overline{U}_1\},\nonumber
\end{eqnarray}
then the new Hamiltonian we construct can be written as:
\begin{equation}
H_0=\omega_1+\omega_2
+\frac{1}{2}\{h_2,\;\overline{U}_1\}.\label{z6}
\end{equation}
It is easy to verify that the eigenfunction set of the new Hamiltonian includes both subsets:
\begin{equation}
h_2\psi^+=\psi^+\; {\rm and}\;  h_2\psi^-=-\psi^-,\label{z7}
\end{equation}
where the subset $\{\psi^+\}$ is identical to the eigenfunction set associated with the original Hamiltonian $H_0$ in Eq. (\ref{z1}).

Now we turn to solve the eigenequation associated with the new Hamiltonian $H_0$, that is:
\begin{equation}
\left(\omega_1+\omega_2
+\frac{1}{2}\{h_2,\;\overline{U}_1\}-E\right)\psi(\boldsymbol{r})=0.\label{z8}
\end{equation}

In the heavy-light quark-antiquark system, we treat the heavy quark as a static source, while the light one is described relativistically by a Dirac spinor. It is easy to verify that $H_0$ commute with all the elements of the standard operator set $\{ {\boldsymbol{j}}^2,j_z,K, S_z \}$ associated with the free Dirac Hamiltonian.
Then the eigenstates of $H_0$ can be labeled by the quantum number set $\{n,j,m_j,k,s\}$ corresponding to the operator set.

The quantum number $k$ can have two opposite values for a eigenstate with quantum number $j$:
\begin{equation}
k=\pm (j+1/2), \; {\rm for}\;\;  l=j\pm 1/2.
\end{equation}
The leading order invariant mass $E^{(0)}$ can be determined by quantum numbers $n$, $j$ and $k$, or equivalently by $n$, $j$ and $l$. The parity of the bound states is determined by $P=(-1)^{l+1}$.

The Dirac spinor with quantum numbers $j$ $m_j$ and $l$ can be written as:
\begin{equation}
\Psi(\boldsymbol{r})
=\left( \begin{array}{cc} g(r)\;y_{j,l_A}^{m_j}(\theta,\varphi)\\i f(r)\;y_{j,l_B}^{m_j}(\theta,\varphi)\end{array}\right ), \label{e35}
\end{equation}
where the subscripts $l_A$ and $l_B$ stand for $l$ and $2j-l$, respectively. The complete expression of $y_{j,l}^{m_j}(\theta,\varphi)$ can be found in Ref. \cite{PE}.

For a bound state of a quark and an antiquark, the wave function will effectively
vanish when the distance between them is large enough. We designate such large typical distance as $L$, then the heavy quark and light antiquark bounded in the meson
can be viewed as restricted in a
limited space, $0 < r < L$. Thus we can expand the radial functions $f(r)$ and $g(r)$ by spherical Bessel functions associated with the distance $L$:
\begin{eqnarray}
g(r)&=&\sum_{i=1}^N\frac{g_i}{N_i^A}j_{l_A}(\frac{a_i^Ar}{L}),\label{e39}\\
f(r)&=&\sum_{\alpha=1}^N\frac{f_\alpha}{N_\alpha^B}j_{l_B}(\frac{a_\alpha^Br}{L}),\label{e40}
\end{eqnarray}
where $N_n$ and $a_n$ are the module and the $n$-th root of the spherical Bessel function $j_l(r)$, respectively.

Inserting Eq. (\ref{e9}) and (\ref{e11}) into Eq. (\ref{z6}), we can rewrite $H_0$ in the matrix form
\begin{eqnarray}
H_0&=&\left( \begin{array}{cc} \omega_1+\omega_2& \\  & \omega_1+\omega_2 \end{array}\right )\nonumber\\
&+&\frac{1}{2}\left( \begin{array}{cc} H_a &H_b \\ H_c & H_d \end{array}\right)+h.c.,\label{e38}
\end{eqnarray}
where $h.c.$ stands for Hermitian conjugate and the operator elements are:
\begin{eqnarray}
H_a&=& \frac{m_2}{\omega_2}(V_v+V_s),\\
H_b&=&\frac{\boldsymbol{\sigma}\cdot\boldsymbol{p}}{\omega_2} (V_v-V_s),\\
H_c&=&\frac{\boldsymbol{\sigma}\cdot\boldsymbol{p}}{\omega_2} (V_v+V_s),\\
H_d&=&\frac{m_2}{\omega_2}(V_s-V_v).
\end{eqnarray}

According to Eqs. (\ref{e39}) and  (\ref{e40}), we can rewrite the eigenequation of $H_0$ in the representation of the state basis constructed by spherical Bessel functions. In this representation, the operator $H_0$ can be written in its matrix form:
\begin{eqnarray}
H_0&=&\left( \begin{array}{cc} <\omega_1+\omega_2>_{ij}& \\  & <\omega_1+\omega_2>_{\alpha\beta} \end{array}\right )\nonumber\\
&+&\frac{1}{2}\left( \begin{array}{cc} <H_a>_{ij}&<H_b>_{i\beta}\\ <H_c>_{\alpha j} & <H_d>_{\alpha\beta} \end{array}\right )+h.c.\label{e41}
\end{eqnarray}

The matrix elements of $H_0$ in the above equation can be calculated by applying the relation
\begin{eqnarray}
(\boldsymbol{\sigma}\cdot\boldsymbol{p}) \;y_{j,l_{\pm}}^{m_j}=\pm i\left(\frac{k\pm1}{r}\pm\frac{d}{dr}\right)
 \;y_{j,l_{\mp}}^{m_j},&&\label{e42}\\
l_{A}=l_{+},  \;l_{B}=l_{-}, &&
\end{eqnarray}
and the eigenequation \cite{ly3}
\begin{equation}
\Omega(p) \; j_l(k r)Y_{lm}(\hat{\boldsymbol{r}})
=\Omega(k) \; j_l(k r)Y_{lm}(\hat{\boldsymbol{r}}),\label{e34}
\end{equation}
where $\Omega(p)$ is a pseudo-differential operator function and $\Omega(k)$ is a normal function, $p$ and $k$ stand for the modules of momentum operator $\boldsymbol{p}$ and momentum $\boldsymbol{k}$, respectively.

With the normalization condition, we can obtain the matrix elements of $H_0$ easily. For the operators with respect to energy of motion, we have
\begin{eqnarray}
<\omega_1(\boldsymbol{p})+\omega_2(\boldsymbol{p})>_{ij}&=&\left[\;\omega_1(\frac{a_i^A}{L})+\omega_2(\frac{a_i^A}{L})\;\right]\delta_{ij},\nonumber\\ \\
<\omega_1(\boldsymbol{p})+\omega_2(\boldsymbol{p})>_{\alpha\beta}&=&\left[\;\omega_1(\frac{a_\alpha^B}{L})+\omega_2(\frac{a_\alpha^B}{L})\;\right]\delta_{\alpha\beta}
.\nonumber \\
\end{eqnarray}
In order to write down the expressions of the elements associated with the interaction potential in a compact form, here we introduce a symbolic notation:
\begin{equation}
\left<\,\hat{O}\,\right>_{m,l_A;n,l_B}=\int_0^L dr\,r^2 \; j_{l_A}(\frac{a_m^Ar}{L})\;\hat{O} \; j_{l_B}(\frac{a_n^Br}{L}),
\end{equation}
then we have:
\begin{eqnarray}
<H_a>_{ij}&=&\frac{1}{N_i^AN_j^A}\frac{m_2}{\omega_2(\frac{a_i^A}{L})}<V_v+V_s>_{i,l_A;j,l_A},\nonumber\\ \\
<H_b>_{i\beta}&=&\frac{1}{N_i^AN_\beta^B}\frac{1}{\omega_2(\frac{a_i^A}{L})}\nonumber\\
&\times&\left<\left(\frac{k-1}{r}-\frac{d}{dr}\right)(V_v-V_s)\right>_{i,l_A;\beta,l_B},\nonumber\\ \\
<H_c>_{\alpha j}&=&\frac{1}{N_\alpha^BN_j^A}\frac{1}{\omega_2(\frac{a_\alpha^B}{L})}\nonumber\\
&\times&\left<\left(\frac{k+1}{r}+\frac{d}{dr}\right)(V_v+V_s)\right>_{\alpha,l_B;j,l_A},\nonumber\\ \\
<H_d>_{\alpha\beta}&=&\frac{1}{N_\alpha^B N_\beta^B}\frac{m_2}{\omega_2(\frac{a_\alpha^B}{L})}<V_s-V_v>_{\alpha,l_B;\beta,l_B}.\nonumber \\
\end{eqnarray}

After calculating every element of the Hermitian matrix of $H_0$, we diagonalize the Hermitian matrix and obtain eigenvalues and eigenvectors. The eigenvalue of the matrix is the eigenenergy of $H_0$, while the eigenvector are associated with the coefficients $g_i,f_\alpha$, which are defined in Eqs. (\ref{e39}) and (\ref{e40}). That is to say, the eigenequation shown in Eq. (\ref{z8}) is solved and the corresponding eigenenergy and eigenfunction are obtained.

Here we turn to discuss the perturbative corrections of $H^\prime$ defined in Eq. (\ref{e27}).
The perturbative term $H^\prime$ does not commute with the standard operators introduced for the free
Dirac Hamiltonian, but still it commutes with the total angular momentum operator $\boldsymbol{J}=\boldsymbol{j}+\boldsymbol{S}$ and the parity operator $\mathcal{P}$ of the bound state.

Thus the quantum number set associated with the total Hamiltonian $H=H_0+H^\prime$ can be denoted by $\{n,J,M_J,P\}$. By using Clebsch-Gordan coefficients, the total wave function of the heavy-light quark-antiquark bound state can be decomposed as follows:
\begin{eqnarray}
\Psi^{(0)}_{n,k,j;J,M_J}(\boldsymbol{r})&=&\sum_{m_j,s} C^{J,M_J}_{j,m_j;1/2,s} \nonumber \\
&\times&\left( \begin{array}{cc} g_{n,k,j}(r)\;y_{j,l_A}^{m_j}(\theta,\varphi)\\i f_{n,k,j}(r)\;y_{j,l_B}^{m_j}(\theta,\varphi)\end{array}\right )\otimes \chi_s,\label{e65}
\end{eqnarray}
with which the corrections and mixings caused by $H^\prime$ can be calculated perturbatively.
The $1/m_Q$ and $1/m^2_Q$ perturbative terms are given in Eqs. (\ref{z9}$\sim$\ref{z11}).

The properties of the eigenfunctions of $H_0$ are of great help in calculating the perturbative corrections. We've already used $h_2\rightarrow1$ to get rid of $h_2$'s at the ends of the perturbative terms. As for the $h_2$'s sandwiched in $H_b^\prime$, $h_2\rightarrow\pm1$ can be applied due to Eq. (\ref{z7}). $H_b^\prime$ can be rewritten as:
\begin{eqnarray}
H_b^{\prime}&=&\frac{1}{4E}\widetilde{U}_2\frac{1}{2}(1-h_2)\left(\widetilde{U}_1\frac{{\boldsymbol{\sigma}}^{(1)}\cdot\boldsymbol{p}}{m_1}
-\frac{{\boldsymbol{\sigma}}^{(1)}\cdot\boldsymbol{p}}{m_1}\overline{U}_1\right) \nonumber\\
&+&h.c.\label{z16}
\end{eqnarray}
Here we define two operators:
\begin{eqnarray}
\hat{A}&=&\widetilde{U}_2, \nonumber\\
\hat{B}&=&\widetilde{U}_1({\boldsymbol{\sigma}}^{(1)}\cdot\boldsymbol{p})
-({\boldsymbol{\sigma}}^{(1)}\cdot\boldsymbol{p})\overline{U}_1.\nonumber
\end{eqnarray}
Then we have:
\begin{equation}
H_b^{\prime}=\frac{1}{4m_1E}\hat{A}\frac{1}{2}(1-h_2)\hat{B}+h.c.\label{z18}
\end{equation}

As discussed at the beginning of this section, the eigenfunction set of $H_0$ can be divided into two parts $\{\psi^+,\,\psi^-\}$, where $\psi^+$ and $\psi^-$ represent the physical and unphysical states, respectively.
Inserting the identity operator composed by the complete set of $H_0$ in Eq. (\ref{z18}), we obtain:
\begin{eqnarray}
H_b^{\prime}&=&\frac{1}{4m_1E}\hat{A}\frac{1}{2}(1-h_2)\sum_{i}|\psi_i><\psi_i|\hat{B}+h.c.\nonumber\\
&=&\frac{1}{4m_1E}\sum_{m}\hat{A}|\psi^-_m><\psi^-_m|\hat{B}+h.c. \label{z19}
\end{eqnarray}

The correction in first order perturbation can be written as:
\begin{equation}
E_{n,l,j,J}=E^{(0)}_{n,l,j}+\delta E^{(1)}_{n,l,j,J}+\delta E^{(2)}_{n,l,j,J},
\end{equation}
where
\begin{equation}
\delta E^{(2)}_{n,l,j,J}=\delta E^{(a)}_{n,l,j,J}+\delta E^{(b)}_{n,l,j,J}.
\end{equation}
From Eq. (\ref{z19}), the correction of $H_b^{\prime}$ for $\psi^+_n$ can be written as:
\begin{eqnarray}
\delta E^{(b)}_{n,l,j,J}&=&\frac{1}{2m_1E_{n,l,j}^{(0)}}\sum_m <\psi^+_n|\hat{A}|\psi^-_m><\psi^-_m|\hat{B}|\psi^+_n>\nonumber\\
&=&\frac{1}{2m_1E_{n,l,j}^{(0)}}\sum_m A_{nm}B_{mn}.
\end{eqnarray}

With the eigenfunctions we obtain, the $1/m_Q$ and $1/m^2_Q$ corrections can be calculated. Then the masses of all the different $J^P$ states are determined.

\section*{IV Numerical results and discussions}

The vector and scalar potentials are chosen to have a Coulombic behavior at short distance and a linear confining behavior at long distance, they can be written in a simple form:
\begin{eqnarray}
V_v(r)&=&-\frac{4\alpha_s(r)}{3r},\label{z12}\\
V_s(r)&=&b\;r+c.\label{z13}
\end{eqnarray}

\begin{figure}[htb]
\begin{center}
\includegraphics[bb=42 636 266 795, width=8.2 cm,clip]{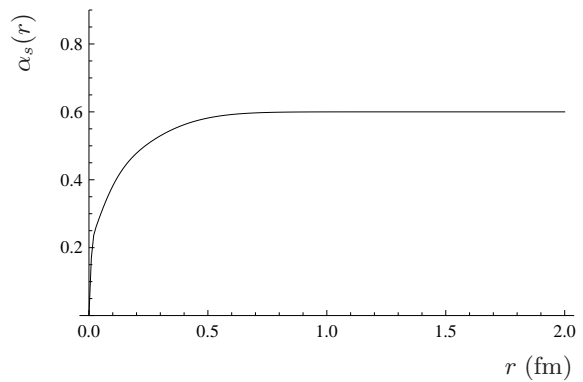}
\caption{The behavior of the running coupling constant $\alpha_s(r)$ with the critical value $\alpha_s^{\rm{critical}}=0.6$. }
\label{as}
\end{center}
\end{figure}

The running coupling constant $\alpha_s(r)$ in the vector potential is derived from the coupling
constant $\alpha_s(Q^2)$ in momentum space via Fourier transformation. It can be parametrized in a more convenient form \cite{GI}:
\begin{equation}
\alpha_s(r)=\sum_i\alpha_i\frac{2}{\sqrt{\pi}}\int_0^{\gamma_i
r}e^{-x^2}dx,
\end{equation}
where $\alpha_i$ and $\gamma_i$ are parameters which can be fitted according to the behavior of the running coupling constant at short distance predicted by QCD. The behavior of $\alpha_s(r)$ is depicted in Fig. \ref{as}. In this work, $\alpha_1=0.25$, $\alpha_2=0.15$, $\alpha_3=0.20$, and $\gamma_1=1/2$, $\gamma_2=\sqrt{10}/2$, $\gamma_3=\sqrt{1000}/2$, the $\alpha_i$ and $\gamma_i$ parameters have the same values as in Ref. \cite{GI}.

\begin{figure}[htb]
\begin{center}
\includegraphics[bb=42 636 266 795, width=8.2 cm,clip]{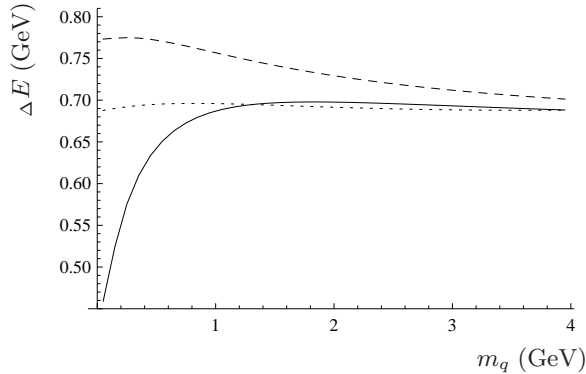}
\caption{The energy gap $\Delta E$ as a function of the light quark mass in $B$ meson. The dashed, dotted and solid lines stand for the Schr\"{o}dinger, Dirac and Bethe-Salpeter formalisms, respectively. }
\label{zfig}
\end{center}
\end{figure}

There are two free parameters in the scalar potential. One is the string tensor constant $b$, which characterizes the confinement of the quark-antiquark system. The other is a phenomenological constant $c$, which is adjusted to give the correct ground state energy level of the heavy-light meson state.
The behavior of the confinement parameters $b$ in this work is quite different from those in usual quark models.
As discussed in our previous paper, the parameter $b$ is responsible for elevating the energy levels of states with higher quantum numbers $n$ or $l$, i.e. the calculated energy gaps between the energy levels of the radial or angular excitations and the ground state depend on the parameter $b$.  But unlike the Dirac Hamiltonian:
\begin{equation}
H_0^{\rm{Dirac}}=\omega_1+ H_2(\boldsymbol{p}) +V_v(r)+ \beta^{(2)}V_s(r),
\end{equation}
or the Schr\"{o}dinger Hamiltonian in Eq. (\ref{e39}), the influence of the confinement potential $V_s(r)$ in Eq. (\ref{z14}) weakens as the mass of light quark decreases. That is to say, in the Bethe-Salpeter formalism the energy level is also sensitive to the light quark mass $m_q$, which is shown in Fig. \ref{zfig} by the energy gap $\Delta E$ between the first radial excitation and the ground state as a function of $m_q$. We take $B$ meson as an example to illustrate the dependence on the quark mass. The values of the parameters, which are fitted for $B$ meson spectrum, are fixed except for the light quark mass of $B$ meson.

From the different shapes of the dashed, dotted and solid lines according to the three schemes, i.e. the Schr\"{o}dinger, Dirac and Bethe-Salpeter formalisms, one can find:
\begin{itemize}
 \item When $m_q$ is taken large enough, the three schemes tend to give the same value for the energy gap $\Delta E$. It indicates the equivalence of the three schemes when dealing with double-heavy mesons.

 \item In the region $m_q < 1 {\rm GeV}$, which is the case for heavy-light mesons, the three schemes give quite different values for the energy gap. It has the pattern:  $\Delta E^{Schr} > \Delta E^{Dirac} > \Delta E^{B-S}$. In order to give the same energy gap for a specific meson, the confinement parameter should be chosen as: $b^{Schr} < b^{Dirac} < b^{B-S}$. The literature supports this sequence. For instance, $b^{Schr}$ is taken as $0.175\;{\rm GeV}^2$ \cite{LY2}, $0.180\;{\rm GeV}^2$ \cite{GI}, $b^{Dirac}$ is taken as $0.257\;{\rm GeV}^2$ \cite{PE}, $0.309\;{\rm GeV}^2$ \cite{MKM}, while $b^{B-S}$ can be taken up to $0.400\;{\rm GeV}^2$ in this work.
 \item In the Schr\"{o}dinger and Dirac schemes, the energy gap changes slowly over $m_q$, this is especially true when $m_q$ is less then 1 Gev, $\Delta E^{Schr}$ and $\Delta E^{Dirac}$ can be viewed as  constants. While in the Bethe-Salpeter scheme, $\Delta E^{B-S}$  changes drastically over $m_q$. From the experimental data, we know that the $\Delta E$'s are not sensitive to their light quark masses. For example, the $\Delta E$'s for both $D$ and $D_s$ mesons are around $0.7\;{\rm GeV}$. Thus $b^{Schr}$ and $b^{Dirac}$ can be taken as a constant, while $b^{B-S}$ varies with the quark mass.
\end{itemize}
Our analysis suggests that in the Bethe-Salpeter formalism  the string tension $b$ depends on the masses of the quark and antiquark, especially the light one of them.

Besides the potential parameters, four quark mass parameters are employed to fit the heavy-light meson spectra. With all the considerations above, our best fitting of the parameters gives the following values
\begin{eqnarray}
&&  m_{u,d}=0.398\; {\rm GeV},\nonumber\\
&&m_s=0.598\;{\rm GeV},\nonumber\\
&&m_c=1.450\; {\rm GeV},\nonumber\\
&& m_b=4.765\; {\rm GeV},\nonumber\\
&&b=\left\{\begin{array}{llll}
 0.390\;{\rm GeV}^2& {\rm for}\; c\bar{q}\; {\rm system}, \\
 0.421\;{\rm GeV}^2& {\rm for}\; b\bar{q}\; {\rm system}, \\
 0.300\;{\rm GeV}^2& {\rm for}\; c\bar{s}\; {\rm system}, \\
 0.316\;{\rm GeV}^2& {\rm for}\; b\bar{s}\; {\rm system},
 \end{array}\right.\nonumber\\
&&c=-0.320\; {\rm GeV}.\nonumber
\end{eqnarray}

In Section III, two numerical parameters $L$ and $N$ are introduced in our calculation. In principle, if the distance $L$ and the size of the expansion basis $N$ are taken to $\infty$, we can obtain the exact solution of the wave equation. Our calculation shows the solution is stable when $L>5$ fm, $N>50$. In this work they are taken as $L=10$ fm, $N=150$, and the size of the matrix of $H_0$ in Eq.(\ref{e41}) is $300\times300$. The Gauss-Legendre quadrature is widely used in our numerical calculation since lots of integrals are involved. After strenuous computation, the meson spectra is obtained. The spectra of the heavy-light $D$, $D_s$, $B$, $B_s$ mesons are fitted based on the data in PDG \cite{PDG}. The numerical results for the spectra of $D$, $D_s$ mesons are presented in Table \ref{t1}, while $B$, $B_s$ mesons are presented in Table \ref{t2}. The calculated spectra are in good agreement with the experimental data. Our results are compared to the results of two other relativistic models \cite{EFG,PE}, one is derived by quasipotential approach, and the other is obtained by reducing the Bethe-Salpeter vertex function.

The result in this work is improved comparing to our previous work \cite{ly3}. Taking the mass difference between the pseudoscalar state and the vector state for example, as shown in Table I, in our previous work, we have:
\begin{eqnarray}
m_{D^*}-m_{D}&=&167\;{\rm GeV}, \nonumber\\
m_{D_s^*}-m_{D_s}&=&161\;{\rm GeV},\nonumber
\end{eqnarray}
while in this work, we have:
\begin{eqnarray}
m_{D^*}-m_{D}&=&137\;{\rm GeV}, \nonumber\\
m_{D_s^*}-m_{D_s}&=&143\;{\rm GeV},\nonumber
\end{eqnarray}
the discrepancy from experimental data is decreased for $D$, $D^*$, $D_s$ and $D_s^*$ states as well as other states.

Theoretical deviations from experimental data mainly occur in the $D_s$ meson sector, specifically, the $D_{s0}^*(2317)$ and $D_{s1}(2460)$ resonances. Our calculations for the two resonances are about $100\;{\rm GeV}$ higher than their masses measured in experiment. The discrepancy may be ascribed to the instantaneous approximation, the naive assumption of the kernel or the $\alpha^2_s(r)$ contributions, i.e. the loop corrections, but it is more likely to be explained beyond naive quark model \cite{ws}. The masses of the two resonances predicted by constituent quark model are generally $100\sim 200$ MeV higher than experiments \cite{JZ,PE,GJ,EGF,LNR}. The mass of $D_0^*$, $2318 \pm 29\;{\rm GeV}$ is almost identical to the mass of $D_{s0}^*$, $2317.8\pm 0.6\;{\rm GeV}$. It cannot be explained in conventional quark model if the difference between the two anomalous resonances in the model is merely their light quark masses $m_s$ and $m_{u,d}$. In this work the confinement parameter $b$ takes different values for different systems, but still it is not capable to explain the small mass difference of the two resonances.

As the $D_{s0}^*$ and $D_{s1}$ lie just below the $DK$ and $D^*K$ threshold, respectively, the authors in Ref. \cite{BCL} have suggested that the two resonance may be $D_{s0}^*(DK)$ and $D_{s1}(D^*K)$ molecular states, while in Refs. \cite{HK1,HK2,BR}, the $D_{s0}^*$ and $D_{s1}$ are considered as $c\bar{s}$ states which are significantly affected by mixing with the $DK$ and $D^*K$ continua. In Ref. \cite{Yu}, the authors suggest that the discrepancy of the calculated masses in quark models can be qualitatively understood as a consequence of self-energy effects due to strong coupled channels.
In Refs. \cite{BEH,EFR,EFFR}, the interpretation of the heavy $J^P (0^+, 1^+)$ spin multiplet as the parity partner of the groundstate $(0^-, 1^-)$ multiplet is proposed. Both theoretical and experimental efforts are required in order to fully understand the nature of the anomalous $D_{s0}^*(2317)$ and $D_{s1}(2460)$ states.

Besides the well-established $1S$ and $1P$  heavy-light meson states, The highly excited states are also calculated in the spectra and the newly observed highly excited meson states beyond $1P$ are identified in our model.

\begin{table*}[!htbp]
\caption{Spectra for $D$ and $D_s$ mesons.  The comparison of the
result in this work with our previous work and other theoretical
results in Refs.\cite{PE,EFG} is presented. All units are in MeV.}
 \label{t1}
\begin{tabular}{C{1.8cm}|C{2cm}|C{4cm}|C{1.8cm}||C{1.8cm}|C{1.8cm}|C{1.8cm}}\hline\hline
$n^jL_J    $ &Meson&$E_\mathrm{expt.}$ \cite{PDG,LHCb}&  $\begin{array}{c}\mathrm{this} \\ \mathrm{work} \end{array}$  &  $\begin{array}{c}\mathrm{previous} \\ \mathrm{work} \end{array}$ \!\!\!\!\cite{ly3}&$\mathrm{Ref.}$  \cite{EFG} &$\mathrm{Ref.}$ \! \cite{PE} \\ \hline\hline
$1^{1/2}S_0$ &  $D$          &$1869.62\pm0.15$  & 1871  & 1859& 1871        &1868        \\
$1^{1/2}S_1$ &  $D^*$        &$2010.28\pm0.13$  & 2008  & 2026& 2010        &2005         \\

$1^{1/2}P_0$ &$D_0^*(2400)^0$&$2318 \pm 29   $  & 2364  & 2357& 2406        &2377          \\
$1^{1/2}P_1$ &               &                  & 2507  & 2529& 2469        &2490           \\
$1^{3/2}P_1$ &$D_1(2420)$    &$2421.3\pm0.6  $  & 2415  & 2434& 2426        &2417          \\
$1^{3/2}P_2$ &$D_2^*(2460)$  &$2464.4 \pm1.9 $  & 2460  & 2482& 2460        &2460          \\

$1^{3/2}D_1$ &               &                  & 2836  & 2852& 2788        &2795         \\
$1^{3/2}D_2$ &               &                  & 2881  & 2900& 2850        &2833         \\
$1^{5/2}D_2$ &$D_J(2740)^0$  &$2737.0\pm 3.5\pm11.2$& 2737  & 2728& 2806    &2775           \\
$1^{5/2}D_3$ &$D_J^*(2760)^0$&$2760.1\pm 1.1\pm3.7$& 2753   & 2753& 2863    &2799          \\

$1^{5/2}F_2$ &               &                  & 3122  & 3107&3090         &3101            \\
$1^{5/2}F_3$ &               &                  & 3139  & 3134&3145         &3123            \\
$1^{7/2}F_3$ &$D^*_J(3000)^0$&$3008.1\pm 4.0$   & 2980  & 2942&3129         &3074            \\

$2^{1/2}S_0$ &$D_J(2580)^0$  &$2579.5\pm 3.4\pm 5.5$& 2594  & 2575& 2581    &2589            \\
$2^{1/2}S_1$ &$D_J^*(2650)^0$&$2649.2\pm 3.5\pm 3.5$& 2672  & 2686& 2632    &2692           \\

$2^{1/2}P_0$ &               &                  & 2895  & 2902&2919         &2949          \\
$2^{1/2}P_1$ &               &                  & 2983  & 2999&3021         &3045          \\
$2^{3/2}P_1$ &               &                  & 2926  & 2932&2932         &2995           \\
$2^{3/2}P_2$ &$D_J(3000)^0$  &$2971.8\pm 8.7$   & 2965  & 2969&3012         &3035           \\

$2^{3/2}D_1$ &               &                  & 3230  & 3228&3228         &           \\
$2^{3/2}D_2$ &               &                  & 3259  & 3260&3307         &           \\
$2^{5/2}D_2$ &               &                  & 3159  & 3139&3259         &            \\
$2^{5/2}D_3$ &               &                  & 3176  & 3160&3335         &            \\

$2^{5/2}F_2$ &               &                  & 3455  & 3425&             &                       \\
$2^{5/2}F_3$ &               &                  & 3465  & 3444&3551         &                          \\
$2^{7/2}F_3$ &               &                  & 3346  & 3301&             &                          \\

\hline\hline

$1^{1/2}S_0$ &$D_s^\pm$      &$1968.49\pm 0.32$ & 1964  & 1949  &1969         & 1965          \\
$1^{1/2}S_1$ &$D_s^{*\pm}$   & $2112.3\pm 0.5$  & 2107  & 2110  &2111          & 2113         \\

$1^{1/2}P_0$ &$D_{s0}^*(2317)$&$2317.8\pm 0.6$  & 2437  & 2412 &2509          & 2487          \\
$1^{1/2}P_1$ &$D_{s1}(2536)$ & $2535.12\pm 0.13$& 2558  & 2562 &2574          & 2605          \\
$1^{3/2}P_1$ &$D_{s1}(2460)$ &$2459.6\pm 0.6$   & 2524  & 2528 & 2536          & 2535         \\
$1^{3/2}P_2$ &$D_{s2}^*(2573)$& $2571.9\pm 0.8$ & 2570  & 2575 &2571           & 2581         \\

$1^{3/2}D_1$ &$D_{s1}^*(2860)^-$&$2859\pm12\pm6\pm23$ \cite{LHCbspin13b}   & 2885  & 2873  &2913         &2913           \\
$1^{3/2}D_2$ &               &                  & 2923  & 2916  & 2961         &2953          \\
$1^{5/2}D_2$ &               &                  & 2857  & 2829  & 2931        &2900           \\
$1^{5/2}D_3$ &$D_{s3}^*(2860)^-$&$2860.5\pm2.6\pm2.5\pm6.0$ \cite{LHCbspin13b}& 2871  & 2852  &2971         &2925           \\

$1^{5/2}F_2$ &               &                  & 3172  & 3128 &3230           &3224          \\
$1^{5/2}F_3$ &               &                  & 3184  & 3152 &3266          & 3247          \\
$1^{7/2}F_3$ &               &                  & 3107  & 3049 &3254           & 3203         \\

$2^{1/2}S_0$ &$D_{sJ}(2632)$& $2632.5\pm 1.7$ \cite{Ds2632}           & 2647  & 2624 &2688           &2700          \\
$2^{1/2}S_1$ &$D_{s1}^*(2710)$ & $2708\pm9^{+11}_{-10}$ \cite{Ds2700}  & 2734  & 2729  &2731          & 2806         \\

$2^{1/2}P_0$ &               &                  & 2945  & 2918  & 3054         &3067          \\
$2^{1/2}P_1$ &$D_{sJ}(3040)$ &$3044\pm8^{+30}_{-5}$ \cite{Ds2009} & 3028  & 3017  & 3154         &3165          \\
$2^{3/2}P_1$ &               &                  & 3009  & 2994 &3067          &3114           \\
$2^{3/2}P_2$ &               &                  & 3047  & 3031  &3142          & 3157         \\

$2^{3/2}D_1$ &               &                  & 3277  & 3247 &3383          &          \\
$2^{3/2}D_2$ &               &                  & 3305  & 3278 &3456          &           \\
$2^{5/2}D_2$ &               &                  & 3260  & 3217 &3403           &          \\
$2^{5/2}D_3$ &               &                  & 3274  & 3237 &3469           &          \\

$2^{5/2}F_2$ &               &                  & 3508  & 3449  &         &           \\
$2^{5/2}F_3$ &               &                  & 3517  & 3468 & 3710         &           \\
$2^{7/2}F_3$ &               &                  & 3459  & 3390 &          &           \\
             &               &                  &       &      &                   &        \\
\hline\hline
\end{tabular}
\end{table*}

\begin{table*}[!htbp]
\caption{Spectra for $B$ and $B_s$  mesons.  The comparison of the
result in this work with our previous work and other theoretical
results in Refs.\cite{PE,EFG} is presented. All units are in MeV.}
 \label{t2}
\begin{tabular}{C{1.8cm}|C{1.8cm}|C{2.8cm}|C{1.8cm}||C{1.8cm}|C{1.8cm}|C{1.8cm}}\hline\hline
$n^jL_J    $ &Meson&$E_\mathrm{expt.}$ \cite{PDG}&  $\begin{array}{c}\mathrm{this} \\ \mathrm{work} \end{array}$  &  $\begin{array}{c}\mathrm{previous} \\ \mathrm{work} \end{array}$ \!\!\!\!\cite{ly3}&$\mathrm{Ref.}$  \cite{EFG} &$\mathrm{Ref.}$ \! \cite{PE} \\ \hline\hline
$1^{1/2}S_0$ &  $B$           &$5279.25\pm 0.17$& 5273  & 5262&5280               &5279                    \\
$1^{1/2}S_1$ &  $B^*$         &$5325.2\pm 0.4$  & 5329  & 5330&5326               &5324                    \\

$1^{1/2}P_0$ &               &                  & 5776  & 5740&5749               &5706                    \\
$1^{1/2}P_1$ &               &                  & 5837  & 5812&5774               &5742                    \\
$1^{3/2}P_1$ &$B_1(5721)$    &$5723.5\pm 2.0$   & 5719  & 5736&5723               &5700                   \\
$1^{3/2}P_2$ &$B_2^*(5747)$  & $5743\pm 5$      & 5739  & 5754&5741               &5714                   \\

$1^{3/2}D_1$ &               &                  & 6143  & 6128&6119               &6025                    \\
$1^{3/2}D_2$ &               &                  & 6165  & 6147&6121               &6037                    \\
$1^{5/2}D_2$ &               &                  & 5993  & 5989&6103               &5985                    \\
$1^{5/2}D_3$ &               &                  & 6004  & 5998&6091               &5993                    \\

$1^{5/2}F_2$ &               &                  & 6379  & 6344&6412               &6264                    \\
$1^{5/2}F_3$ &               &                  & 6391  & 6354&6420               &6271                  \\
$1^{7/2}F_3$ &               &                  & 6202  & 6175&6391               &6220                     \\

$2^{1/2}S_0$ &               &                  & 5957  & 5915&5890               &5886                     \\
$2^{1/2}S_1$ &               &                  & 5997  & 5959&5906               &5920                    \\

$2^{1/2}P_0$ &               &                  & 6270  & 6211&6221               &6163                     \\
$2^{1/2}P_1$ &               &                  & 6301  & 6249&6281               &6194                    \\
$2^{3/2}P_1$ &               &                  & 6216  & 6189&6209               &6175                    \\
$2^{3/2}P_2$ &               &                  & 6232  & 6200&6260               &6188                     \\

$2^{3/2}D_1$ &               &                  & 6514  & 6458&6534               &                    \\
$2^{3/2}D_2$ &               &                  & 6527  & 6471&6554               &                     \\
$2^{5/2}D_2$ &               &                  & 6401  & 6357&6528               &                     \\
$2^{5/2}D_3$ &               &                  & 6411  & 6365&6542               &                   \\

$2^{5/2}F_2$ &               &                  & 6692  & 6621&                   &                     \\
$2^{5/2}F_3$ &               &                  & 6700  & 6629&6786               &                    \\
$2^{7/2}F_3$ &               &                  & 6553  & 6493&                   &                    \\
\hline\hline

$1^{1/2}S_0$ &$B_s$          &$5366.77\pm 0.24$      & 5363  & 5337&5372            &5373          \\
$1^{1/2}S_1$ &$B_s^*$        &$5415.4^{+2.4}_{-2.1} $& 5419  & 5405&5414            &5421          \\

$1^{1/2}P_0$ &               &                  & 5811  & 5776  &5833               &5804        \\
$1^{1/2}P_1$ &               &                  & 5864  & 5841  &5865               &5842        \\
$1^{3/2}P_1$ &$B_{s1}(5830)$ &$5829.4\pm 0.7$   & 5819  & 5824  &5831               &5805        \\
$1^{3/2}P_2$ &$B_{s2}^*(5840)$&$5839.7\pm 0.6$  & 5838  & 5843  &5842               &5820        \\

$1^{3/2}D_1$ &               &                  & 6167  & 6146  &6209               &6127        \\
$1^{3/2}D_2$ &               &                  & 6186  & 6163  &6218               &6140        \\
$1^{5/2}D_2$ &               &                  & 6098  & 6085  &6189               &6095        \\
$1^{5/2}D_3$ &               &                  & 6109  & 6094  &6191               &6103        \\

$1^{5/2}F_2$ &               &                  & 6405  & 6363  &6501               &6369        \\
$1^{5/2}F_3$ &               &                  & 6416  & 6373  &6515               &6376        \\
$1^{7/2}F_3$ &               &                  & 6313  & 6276  &6468               &6332        \\

$2^{1/2}S_0$ &               &                  & 6010  & 5961  &5976               &5985        \\
$2^{1/2}S_1$ &               &                  & 6048  & 6003  &5992               &6019        \\

$2^{1/2}P_0$ &               &                  & 6291  & 6227  &6318               &6264        \\
$2^{1/2}P_1$ &               &                  & 6323  & 6266  &6345               &6296        \\
$2^{3/2}P_1$ &               &                  & 6288  & 6249  &6321               &6278        \\
$2^{3/2}P_2$ &               &                  & 6304  & 6263  &6359               &6292        \\

$2^{3/2}D_1$ &               &                  & 6540  & 6478  &6629               &        \\
$2^{3/2}D_2$ &               &                  & 6553  & 6491  &6651               &        \\
$2^{5/2}D_2$ &               &                  & 6487  & 6434  &6625               &        \\
$2^{5/2}D_3$ &               &                  & 6496  & 6441  &6637               &        \\

$2^{5/2}F_2$ &               &                  & 6723  & 6647  &                   &        \\
$2^{5/2}F_3$ &               &                  & 6731  & 6654  &6880               &        \\
$2^{7/2}F_3$ &               &                  & 6650  & 6580  &                   &        \\
             &               &                  &       &       &                   &        \\
\hline\hline
\end{tabular}
\end{table*}

As for $D$ mesons, in the mass region $2500 \sim 3000$ MeV several resonances are measured by LHCb collaboration \cite{LHCb}.
The assignments of these states are listed in the upper part of Table I, where the resonances $D_J(2740)$, $D_J^*(2760)$, $D_J(3000)$ are identified as $n=1$ states and the resonances $D_J(2580)$, $D_J^*(2650)$, $D_J^*(3000)$ are identified as
radially exited states with $n=2$.
In our predicted spectrum for $D$ meson, $D_J(2740)$ and $D_J^*(2760)$ are identified as the $|1^{5/2}D_2\rangle$ state with $J=2^-$ and the $|1^{5/2}D_3\rangle$ state with $J=3^-$, respectively.  Our best assignment for $D_J^* (3000)$ is the $|1^{7/2}F_3\rangle$ state and $D_J(3000)$  the $|2^{3/2}P_2\rangle$ state, although in Ref. \cite{LHCb} they favor the natural and unnatural parity, respectively. The last two resonances $D_J(2580)$ and $D_J^*(2650)$ are identified as the first radial excitations of the ground $D$ and $D^*$ states. Recently, LHCb collaboration observed $D_J^*(2650)$ and $D_J^*(2760)$, their masses and widths were measured as \cite{LHCb2016}:
\begin{eqnarray}
M(D_{1}^*(2680)^0)&=&2681.1\pm5.6\pm4.9\pm13.1\;\mbox{MeV},\nonumber\\
\Gamma(D_{1}^*(2680)^0)&=&186.7\pm8.5\pm8.6\pm8.2\;\mbox{MeV},\nonumber\\
M(D_{3}^*(2760)^0)&=&2775.5\pm4.5\pm4.5\pm4.7\;\mbox{MeV},\nonumber\\
\Gamma(D_{3}^*(2760)^0)&=&95.3\pm9.6\pm7.9\pm33.1\;\mbox{MeV}.\nonumber
\end{eqnarray}
From Table \ref{t1}, one can see our results favor the measurement.

As for $D_s$ mesons, several states beyond $1P$ state have been observed, their masses and identifications are presented in the lower part of Table \ref{t1}.  Recently, LHCb collaboration identifies $D_{sJ}^*(2860)$ as an admixture of two resonances: $D_{s3}^*(2860)^-$ and $D_{s1}^*(2860)^-$ \cite{LHCbspin13b,LHCbspin13a}, with their masses measured as $2859\pm12\pm6\pm23\;\mbox{MeV}$ and $2860.5\pm2.6\pm2.5\pm6.0\;\mbox{MeV}$, respectively. In Refs. \cite{PE,EFG} cited in Table \ref{t1}, their predictions do not favor this identification, with their calculations generally 60 MeV higher than the measured masses. While our results for both $|1^{3/2}D_1\rangle$ and $|1^{5/2}D_3\rangle$ are around 2860 MeV, the two resonances can be interpreted as members of the 1D family with $J^P=1^-$ and $3^-$.
The resonances $D_{sJ}(2632)$ , $D_{s1}^*(2710)$ and $D_{sJ}(3040)$ are identified as radially exited states with $n=2$ in our model. The $D_{sJ}(2632)$ was firstly observed by SELEX Collaboration at a mass of $2632.5\pm1.7$  MeV, it can be assigned as the $|2^{1/2}S_0\rangle$. The assignment for $D_{s1}^*(2710)$ is proposed as $J^P=1^-$  in Refs. \cite{CFNR,CTLS}, which agree with our prediction as our calculated mass for $|2^{1/2}S_1\rangle$ is close to its experimental mass $2708\pm9^{+11}_{-10}$ MeV \cite{Ds2700}. The $D_{sJ}(3040)$ resonance is observed in the $D^*K$ mass spectrum at a mass of $3044\pm8_{-5}^{+30}$ MeV by BABAR Collaboration \cite{Ds2009}. Here we assign it as $|2^{1/2}P_1\rangle$ in our predicted $D_s$ meson spectrum.

 In the b-flavored meson sector, experimental data for excited $B$ meson states are limited for now. But still several  b-flavored mesons are observed \cite{CCL}.
The strangeless resonances $B_J(5840)^0$ and $B(5970)^0$ were measured by the LHCb and CDF Collaborations, respectively \cite{LHCba,CDFa}. The stranged $B^*_{sJ}(5850)$ were observed by the OPAL Collaboration \cite{OPAL}. Their masses were measured as:
\begin{eqnarray}
M(B_J(5840))&=& 5862.9\pm5.0\pm6.7\pm0.2\;\mbox{MeV},\nonumber\\
M(B(5970)^0)&=& 5978\pm5\pm12\;\mbox{MeV},\nonumber\\
M(B^*_{sJ}(5850))&=& 5853\pm15\;\mbox{MeV}.\nonumber
\end{eqnarray}
In table \ref{t2}, we can identify $B_J(5840)^0$ and $B(5970)^0$ as $|1^{1/2}P_1\rangle$ and  $|2^{1/2}S_1\rangle$, respectively in the spectrum of $B$ meson, while  $B^*_{sJ}(5850)$ can be assigned as $|1^{1/2}P_1\rangle$ in the spectrum of $B_s$ meson.

\begin{figure*}
\centering
\scalebox{1}{\epsfig{file=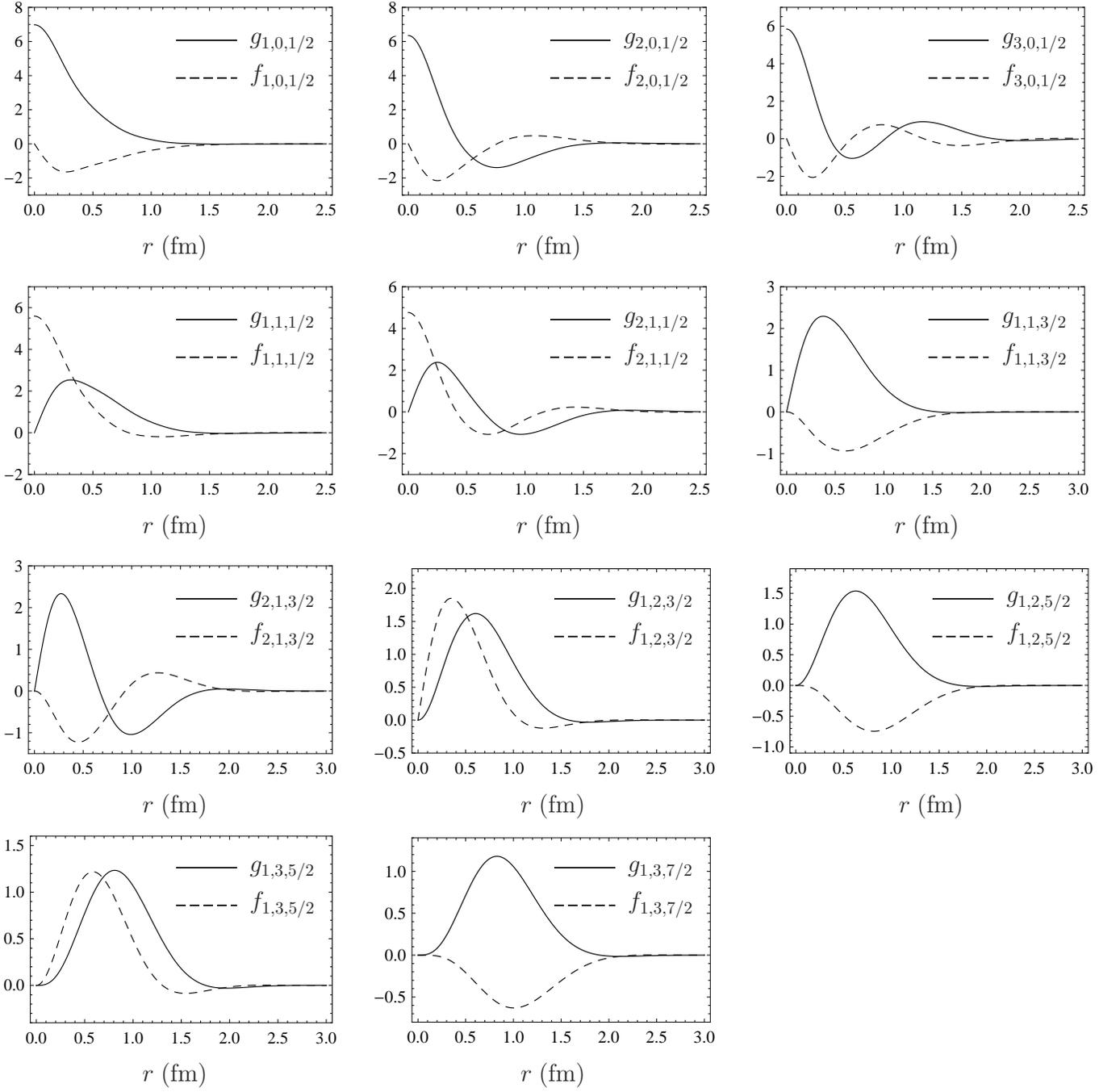}}
\caption{The radial wave functions $g_{n,l,j}(r)$ and $f_{n,l,j}(r)$ for physical $D$ meson states as an example. The wave functions are the radial part of the solution of the eigenequation associated with $H_0$.}
\label{f1}
\end{figure*}

\begin{figure*}
\centering
\scalebox{1}{\epsfig{file=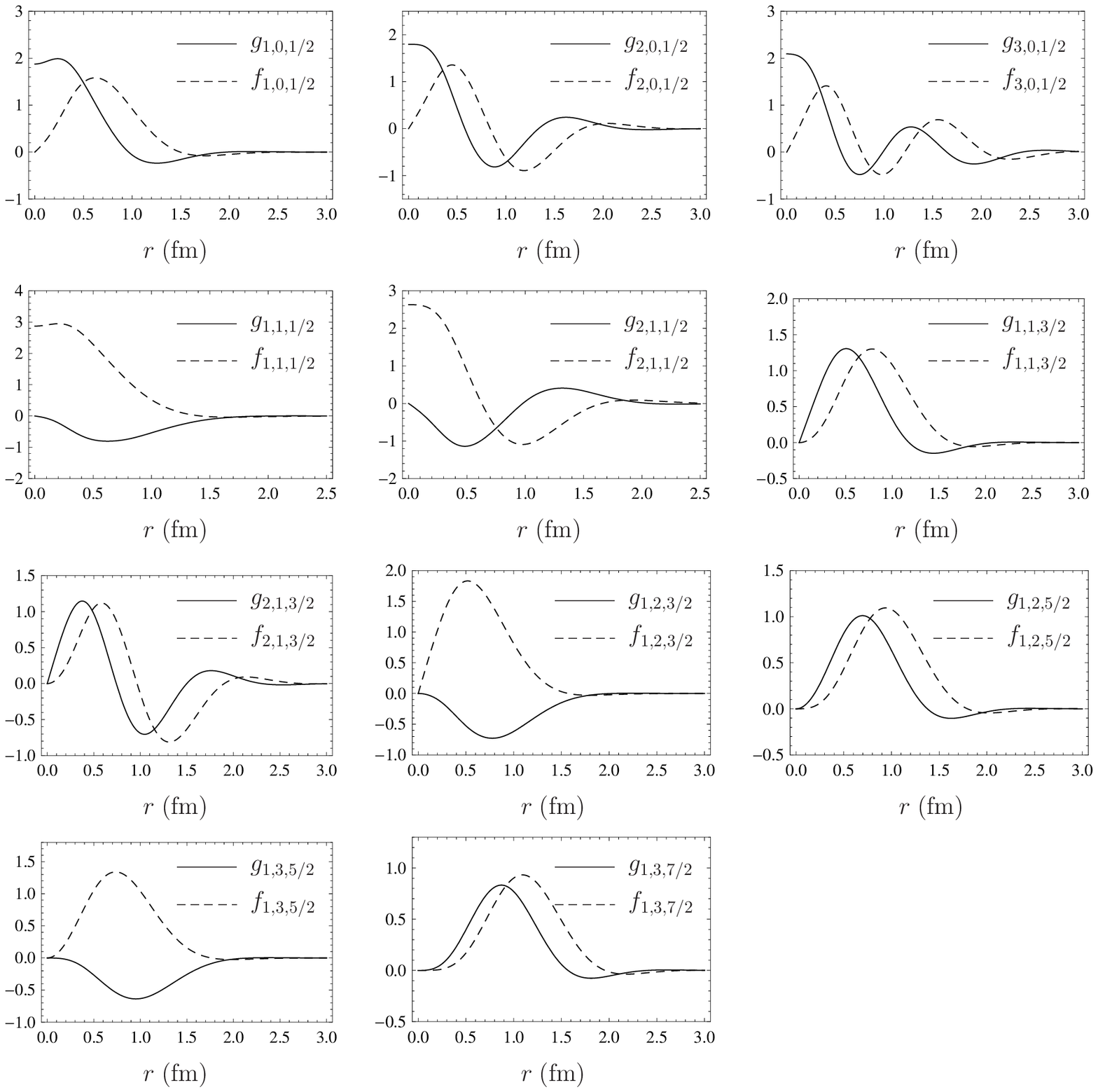}}
\caption{The radial wave functions $g_{n,l,j}(r)$ and $f_{n,l,j}(r)$ for unphysical $D$ meson states as an example. The wave functions are the radial part of the solution of the eigenequation associated with $H_0$.}
\label{f2}
\end{figure*}

Finally, after solving the wave equation one can  obtain not only the eigenenergy of each bound state but also their wave functions. The radial wave functions $g_{n,l,j}(r)$ and $f_{n,l,j}(r)$ for physical and unphysical $D$ meson states are depicted as an example in Fig. \ref{f1} and \ref{f2}, respectively. We stress that the solution of the eigenequation associated with the original $H_0$  in Eq. (\ref{z14}) gives only the wavefunctions of physical states depicted in Fig. \ref{f1}. In Section III we construct a new $H_0$ for heavy-light systems in Eq. (\ref{z6}), the unphysical states depicted in Fig. \ref{f2} is due to the new $H_0$ for which the original one is substituted.

\section*{V Summary}

The spectra of heavy-light mesons are restudied in a relativistic  model, which is derived by reducing the instantaneous Bethe-Salpeter equation. The kernel is chosen to be the standard combination of linear scalar and Coulombic vector. By applying the Foldy-Wouthuysen transformation on the heavy quark, the Hamiltonian for heavy-light quark-antiquark system is calculated up to order $1/m_Q^2$.
We find that in the framework of instantaneous Bethe-Salpeter equation the string tension $b$ in the confinement potential is sensitive to the masses of the constituent quarks in the meson.
The spectra of $D$, $D_s$, $B$ and $B_s$ mesons calculated in the relativistic model. Most of the heavy-light meson states can be accommodated successfully in our model except for the anomalous $D_{s0}^*(2317)$ and $D_{s1}(2460)$ resonances. In the Bethe-Salpeter formalism, the assumption of the interaction kernel for mesons is rather a priori, kernels with other spin structures can also be studied.
In this work, we only restrict in calculating the spectra of heavy light mesons. With the wave functions obtained when solving the wave equation, $B$ and $D$ decays can be studied in further researches.


\section*{Acknowledgments}
This work is supported in part by the National Natural Science Foundation of China (Grants  No. 11375208, No. 11521505, No. 11235005, and No. 11621131001).



\end{document}